# Evolving basal slip under glaciers and ice streams

Anders Damsgaard[a,b], Jenny Suckale[a,c,d,1], and Liran Goren[e]

a: Department of Geophysics, Stanford University, 397 Panama Mall, Stanford, CA 94305, USA. b: Department of Geoscience, Aarhus University, Høegh-Guldbergs Gade 2, 8000 Aarhus C, Denmark. c: Institute for Computational and Mathematical Engineering, Stanford University, 475 Via Ortega, Stanford, CA 94305, USA. d: Department of Civil and Environmental Engineering, Stanford University, 473 Via Ortega, Stanford, CA 94305, USA. e: Department of Geological and Environmental Sciences, Ben-Gurion University of the Negev, P.O.B. 653, Beer-Sheva 84105, Israel. 1: E-mail: jsuckale@stanford.edu

**Many fast-flowing glaciers and ice streams move over beds consisting of reworked sediments and erosional products, commonly referred to as till. The complex interplay between ice, meltwater, and till at the subglacial bed connects several fundamental problems in glaciology including the debate about rapid mass loss from the ice sheets, the formation and evolution of subglacial landforms, and the storage and transport of subglacial water. In-situ measurements have probed the subglacial bed, but provide surprisingly variable and seemingly inconsistent evidence of the depth where deformation occurs, even at a given field site. These observations suggest that subglacial beds are inherently dynamic. The goal of this paper is to advance our understanding of the physical processes that contribute to the dynamics of subglacial beds as reflected in existing observations of basal deformation. We build on recent advances in modeling dense, granular flows to derive a new numerical model for water-saturated till. Our model demonstrates that changes in the force balance or temporal variations in water pressure can shift slip away from the ice-bed interface and far into the bed, causing episodes of significantly enhanced till transport because the entire till layer above the deep slip interface becomes mobilized. We compare our model results against observations of basal deformation from both mountain glacier and ice-stream settings in the past and present. We also present an analytical solution to assess the variability of till transport for different glacial settings and hydraulic properties.**

**Significance statement**
Many glaciers and ice streams move rapidly because they slip over till deposits saturated with meltwater. The dynamic interplay between ice and till reshapes the bed, creating landforms preserved from past glaciations. Understanding the processes that shaped these landforms allows us to leverage the imprint left by past glaciations as constraints for modeling current and future deglaciation. However, observations provide seemingly inconsistent evidence of basal processes even at the same field site. We argue that the observed variability reflects different dynamic regimes in the subglacial environment. Our result suggests that the till-ice interface is highly dynamic with till deformation shifting from shallow to deep with changes in the force balance or variable water influx.

Keywords: Ice sheet dynamics | Till deformation | Subglacial hydrology | Dense granular flows | Basal sliding

The Intergovernmental Panel on Climate Change (IPCC) ascribes the largest uncertainty in future sea level rise to the ice dynamics of our two ice sheets, Greenland and Antarctica (1). Most of this uncertainty arises from specifying meaningful boundary conditions at both the base of the land ice and its seaward margins (2). While ice-ocean interactions have received considerable attention over recent years (3–5), our ability to predict the evolution of basal conditions underneath the ice sheets remains limited.

The debate over the evolution of basal conditions during deglaciation is closely tied to the question of where basal slip is accommodated. The fast speed of the ice leaves little doubt that motion is facilitated primarily by localized basal slip rather than distributed deformation inside the ice, but observations of where slip occurs provide conflicting evidence. Field observations range from basal slip occurring almost exclusively at the ice-bed interface (6, 7), to partial slip at the ice-bed interface combined with shallow deformation within the upper portion of the till (8, 9), and deep basal slip meters below the ice-bed interface (10, 11). Interestingly, the depth of basal slip can vary by several meters even at the same field site as in the case of Black Rapids Glacier, Alaska (9, 10). Observations hence suggest that the depth of basal slip is a dynamic variable that evolves in time.

The depth of basal slip determines both the rate and the volume of till transport, and directly influences the pace at which the glacier reshapes its bed (19–22). The bedforms emerging from this reshaping process alter ice motion and till deformation by locally changing the bed slopes, rerouting subglacial water (23–25) and, potentially, creating kinematic barriers to ice transport. Leveraging the information imprinted in bedforms preserved from past glaciations (26, 27) to inform models of current and future deglaciation, requires understanding the feedback between ice motion, till deformation and water storage and drainage.

The goal of this paper is to advance our understanding of the physical processes that contribute to the dynamics of subglacial beds as reflected in existing, seemingly inconsistent observations of basal deformation. We hypothesize that the variability in observational evidence reflects different dynamic regimes arising from the complex interplay between ice, meltwater and till in the subglacial environment. We test this hypothesis by deriving a continuum model for water-saturated subglacial till that is consistent with laboratory measurements of a near-plastic till rheology (13–16). Our model builds on recent advances in continuum descriptions of dense, granular behavior by Henann and Kamrin (28) and integrates it with progress in understanding the coupling of water flux through granular media at the scale of individual grains (29–31).

Over the past two decades, many granular mechanics models have shifted towards statistical approaches (32) and away from classical solid mechanics (33). The main advance provided by this line of thinking is the ability to represent non-local effects, which represent deformation in places where the granular medium is not strictly at the yield stress due to interactions with failing patches nearby. Granular shear zones are an example of non-locality (34–36). In the non-local granular fluidity (NGF) model by Henann and Kamrin (28), a *fluidity* field variable accounts for the non-local effects on deformation. Microscopically, fluidity is related to velocity fluctuations in the granular medium (37). The spatial effect of non-locality scales with a dimensionless material dependent parameter $A$, and exhibits a weak rate dependence on friction beyond yield scales with dimensionless parameter $b$.

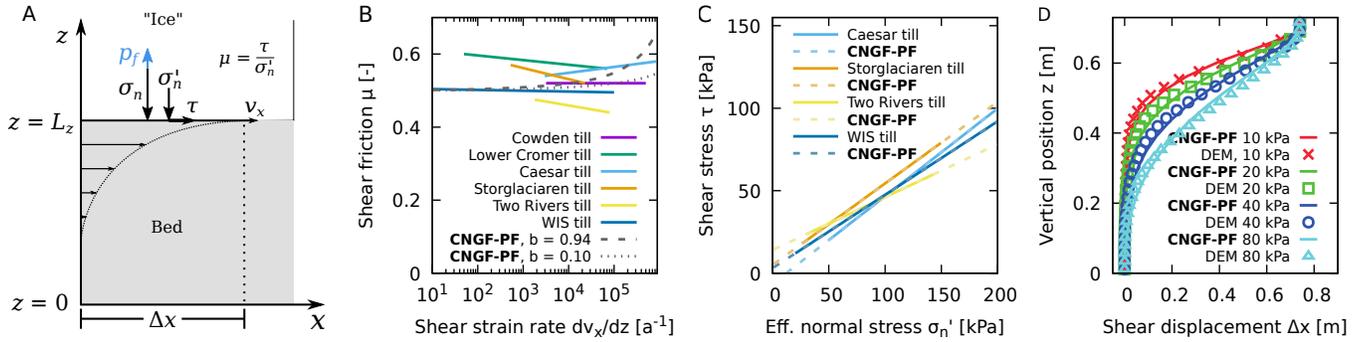

**Fig. 1.** Comparison between shear experiments on subglacial till and CNGF-PF continuum model simulations of till behavior. A) CNGF-PF model setup in one-dimensional shear. The upper boundary is constant speed $v_x$ (speed controlled), or constant shear stress $\tau = \mu \sigma_n^r$ (stress controlled). B) Rate dependence of critical-state friction in laboratory experiments on till (12–15), after (16), and in simulations with material friction $\mu_s$ = 0.5 and effective normal stress $\sigma_n^r$ = 100 kPa. The dimensionless parameter $b$ controls frictional rate dependence (Eq. 5), where $b$ = 0.94 is appropriate for glass beads (17). C) Mohr-Coulomb analysis of actual till samples (13–15) in comparison to simulations. D) Modeled strain distribution under varying effective normal stress ($\sigma_n^r$) with the discrete-element method (DEM, (18)) and the CNGF-PF model.

The original non-local fluidity model accurately describes the strain distribution in a variety of experimental settings (28), but it assumes a dry, cohesionless granular medium. In the subglacial context, in-situ measurements document significant temporal variations in both water pressure and shear strain rates (6, 8, 10, 38). These variations result from both internal till dynamics (39) and external water input (40), highlighting that meltwater is a crucial component of the overall dynamics. In the current study, we capture the interplay between till strength and water percolation by incorporating effective stress and pore-pressure diffusion that depends on porosity and permeability (29, 30). We also consider cohesion in subglacial tills, which tends to increase with clay content (41), by adding a strength contribution from cohesion to the model. We refer to our model as the cohesive non-local granular fluidity model with pore fluid (CNGF-PF).

To test the ability of our model to reproduce the range of observed behavior in the field, we consider two idealized cases, a speed-controlled and a stress-controlled ice-bed interface (see Fig.1A). Speed-controlled conditions are typical for laboratory shear experiments, and represent a subglacial setting where changes in bed friction do not immediately alter the ice flow velocity because the force balance includes other prominent terms such as topography or shear stress in the lateral margins. Specifically, the interfacial shear velocity at the base may vary over time, but is not dependent on the local basal resistance contributed by subglacial friction. An example of this limit is Storglaciären Glacier, Sweden, where subglacial strain rates and surface velocities are largely uncorrelated, at least at the place and time of measurement by Hooke et al. (8). Similarly, the slip speed of several outlet glaciers in Greenland show little sensitivity to variations in meltwater production (42). In contrast, stress-controlled simulations approximate conditions where ice flow velocity responds to changes in subglacial strain rate directly. Whillans Ice Stream, West Antarctica is an example of this setting, where a low surface slope and low driving stress results in stick-slip movement (43). Evidently, most real glacier settings fall along the spectrum spanned between the two limits, depending on how important basal friction is to the overall stress balance.

## Results

### Consistency with laboratory measurements and granular-scale simulations

Natural glaciers systems show extremely variable strain rates of up to $\sim 5 \times 10^3$ a$^{-1}$. When studied over five orders of strain-rate magnitude, some tested natural tills show slight rate weakening, while others are slightly rate strengthening (Fig. 1B). Much like the natural tills, our simulated till is effectively rate-independent over most of the range. At extreme shear-strain rates (Fig. 1B), higher values of $b$ provide larger frictional resistance, leading to a slight hardening. Our model exactly reproduces the Mohr-Coulomb behavior of real tills (13–15) when we insert published values for internal friction $\mu_s$ and cohesion $C$ into the model (Fig. 1C).

The dimensionless non-locality parameter $A$ cannot be constrained against laboratory data on tills, because we are not aware of any experiments that analyze the effect of normal stress on the strain distribution in till. Instead, we compare the modeled strain distribution with discrete-element method results from (18). By inserting relevant material parameters for grain size, friction, stress, and shear velocity (DEM, Table S1), our model approximates the strain distribution well when $A$ = 0.5 (Fig. 1D). Both models show that sediment transport is pressure dependent, with low effective normal stresses producing shallow deformation, and high effective normal stresses deepening the material mobilization. However, the DEM results required more than two months of computational time on a modern GPU cluster, whereas the continuum model is completed in a fraction of a second, albeit without details of individual particle kinematics and dynamical adjustment towards the critical state. A local-only granular model (Fig. S1) cannot produce a good fit to the DEM results, which highlights the importance of non-locality to accurately describe shear strain.

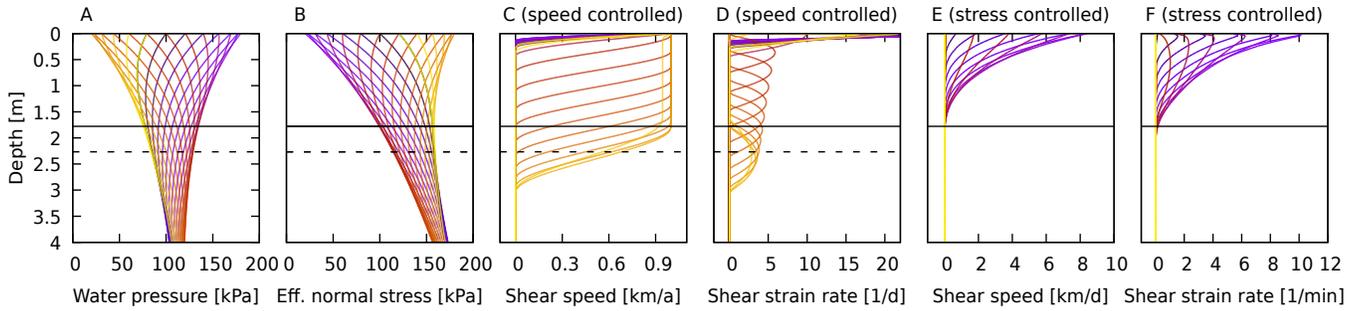

**Fig. 2.** Pore-pressure diffusion and strain distribution with depth with a sinusoidal water-pressure forcing from the top. A) The water-pressure forcing has a daily periodicity, and modulates the effective normal stress (B) with exponential decay at depth. C and D show the speed and shear-strain rate distribution with depth for shear under constant speed, while E and F show the same parameters for shear in stress-controlled configuration. Plot lines are one hour apart in simulation time and lines are colored according to time. The full horizontal line marks the skin depth from Eq. 1, and the dashed line marks the expected maximum deformation depth from Eq. 2.

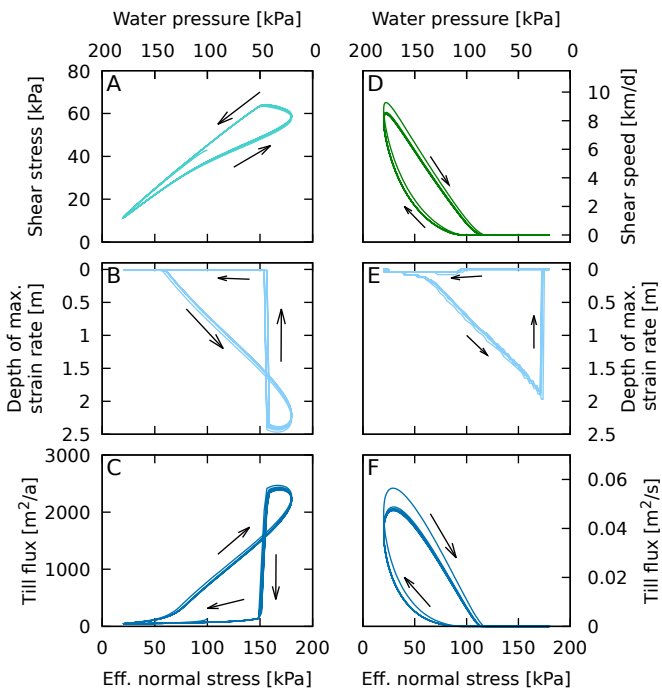

**Fig. 3.** Deformation dynamics under sinusoidal forcing in water pressure and effective normal stress at the ice-bed interface. Panels A-C are under speed-controlled shear, D-F) are under stress-controlled shear. Black arrows in panels A-F denote the temporal evolution.

## Meltwater pulses can lead to deep slip in speed-controlled settings

To test the hypothesis that meltwater input can alter the depth at which basal slip is concentrated, we impose daily sinusoidal variations in water pressure at the top of the model domain and simulate the resultant shear dynamics over seven days (full time series in Fig. S2). The effective normal stress generally increases with depth due to increasing grain overburden. However, the time-variable input in meltwater at the ice-bed interface and its downward diffusion modulate this depth-trend. Due to diffusion, water pressure perturbations decay exponentially with depth, and travel with a phase shift (Fig. 2A). As a consequence, the distribution of water pressure can create minima in effective normal stress at significant depth below the ice-bed interface, such that the effective stress profile locally reverses (Fig. 2B).

Under speed-controlled conditions (Fig. 2C,D), simulations show highly time-variable deformation patterns throughout the till. Granular failure and maximum shear deformation occur where the effective normal stress is at its minimum, resulting in a plug-like flow during reversal of the effective stress depthtrend. This occurs predominantly when the water pressure at the ice-bed interface is close to its minimum and during its rising limb. For a brief period of time, active shear occurs at two depths, following two local minima in the effective stress trend, and deep deformation is accompanied by weak shear at the surface.

Under the same water-pressure forcing, the stress-controlled conditions (Fig. 2E,F) produce a very different dynamical pattern of deformation. Shear occurs almost exclusively in the uppermost part of the modeled till layer. The cyclic effective stress in combination with the frictional properties of the till lead to stick-slip behavior, which is associated with dramatic variations in the shear speed (Fig. 2E). When water pressure at the top decreases and effective stress increases, shear strain rates decrease from the top and are maximal just below the ice-bed interface (Fig. 1F). When the effective stress at the top further increases, the shear stress is insufficient to overcome even the minimal effective stress in the layer, and till deformation and transport stagnate.

## Till flux is controlled by the coupling at the ice-till interface

In Fig. 3, we replot the simulations from Fig. 2 to clarify the relationship between water pressure and stress, strain rate, and till flux. Under speed-controlled conditions, the shear stress varies in an approximately linear manner as predicted for Mohr-Coulomb materials (33), but with hysteresis at high effective normal stresses (Fig. 3A). Shear stress is higher during deep shear because the lithostatic contribution increases effective normal stress and shear strength at depth. When driven by stress-controlled conditions, the simulated water-sediment system shows stick-slip behavior with locally very high slip speeds and strong hysteresis (Fig. 3D). We emphasize that the specific values for the slip speed and till flux are not realistic for a glacial setting, because our model neglects the far-field stresses like lateral support from the margins. It is also possible that real ice-bed interfaces transition from a stress-controlled to a speed-controlled configuration during slip.

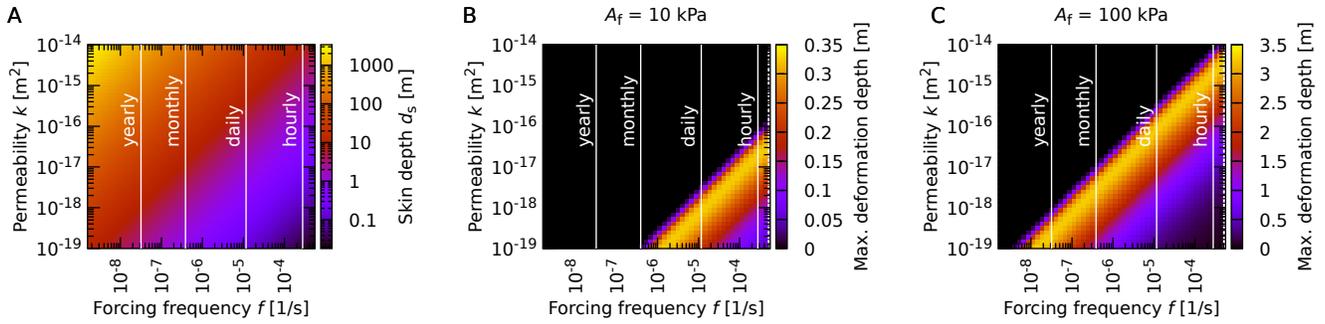

**Fig. 4.** Analytical solutions to skin depth and maximum deformation depth from a sinusoidal water-pressure forcing at the ice-bed interface. A) Skin depth of pore-pressure fluctuations (Eq. 1) with forcing frequencies ranging from yearly to hourly periods. The permeability spans values common for tills. B) Depth of maximum deformation with a water-pressure forcing of $A_f = 10$ kPa. C) Same as (B) but with $A_f = 100$ kPa.

Our simulations show that speed-controlled interfaces favor deep deformation during periods where water-pressure at the ice-bed interface is at its lowest magnitude (Fig. 3B). However, under stress-controlled conditions the deep deformation occurs with insignificant strain rates during the stick phase (Fig. 2E and Fig. 3D,E). The trends in when and how much till transport occurs are therefore drastically different for the two cases. The largest sediment transport under speed-controlled conditions occurs with low water pressures and high effective normal-stresses at the ice-bed interface (Fig. 3B,C). Instead, the majority of till transport under stress-controlled conditions occurs as shallow deformation during rapid slip events and high water pressures (Fig. 3E,F).

## Deep slip depends on water-pressure variations and till permeability

While the previous two sections shed light on the effect of meltwater pulses and the importance of the boundary condition imposed at the till-ice interface, the occurrence of deep slip also depends sensitively on the hydraulic permeability of the till layer which can vary by orders of magnitude. Here, we derive an analytical solution that can provide a quick order-of-magnitude estimate for the depth at which deep slip is expected to occur as a function of till permeability and amplitude and frequency of the water pressure variations.

Our simulations in Figs. 2 and 3 demonstrate that deformation throughout the till thickness follows the minimum in effective normal stress, and reaches its maximum depth at the time when water pressure at the ice-bed interface is at its minimum value. The deepest deformation under speed-controlled conditions hence depends on the diffusion of pore-pressure perturbations away from the ice-bed interface. The length-scale of pressure-perturbation decay due to diffusion is characterized by the skin depth $d_s$ [m]. Assuming that fluid and hydraulic skeleton properties are constant and the layer is sufficiently thick:

$$d_s = \sqrt{\frac{k}{\phi \eta_f \beta_f \pi f}}, \tag{1}$$

where $k$ [m$^2$] is the hydraulic permeability, $\eta_f$ [Pa s] and $\beta_f$ [Pa$^{-1}$] are the dynamic viscosity and the comprehensibility of water, respectively, and $f$ [s$^{-1}$] is the frequency of the pore-pressure variations at the ice-bed interface.

Figure 4A shows the skin depth for water at 0°C under a range of permeabilities and water-pressure forcing frequencies. Importantly, however, the deformation pattern also depends on the distribution of effective stress and hence on the pressure-perturbation amplitude $A_f$ [m] (Fig. S3, S4), which means that the skin depth alone is insufficient to infer whether deep deformation occurs. We derive an analytical solution for diffusive pressure perturbation to find the largest depth $z'$ beneath the ice-bed interface where the effective stress is minimal over the course of a pressure-perturbation cycle (see Supplementary Information Text S2 for derivation). As the shear zone under speed-controlled conditions follows the effective normal stress minimum, this depth corresponds to the deepest expected position of the shear zone midpoint during a cycle of water-pressure forcing at the ice-bed interface:

$$0 = \sqrt{2} \sin\left(\frac{7\pi}{4} - \frac{z'}{d_s}\right) + \frac{(\rho_s - \rho_f)G d_s}{A_f} \exp\left(\frac{z'}{d_s}\right). \tag{2}$$

where $\rho_s$ and $\rho_f$ [kg m$^{-3}$] is the till and water density, respectively, and $G$ [m s$^{-2}$] is the gravitational acceleration. The prediction of the analytical solution matches our simulations well (depth marked by dashed horizontal line in Fig. 2).

We plot solutions to Eq. 2 for various combinations of water-pressure perturbation amplitude ($A_f$), perturbation frequency ($f$), and permeability ($k$) in Fig. 4B and C. For a given pressure-perturbation amplitude, deep deformation does not occur above a threshold value for the skin depth. Beneath this threshold, deformation quickly deepens to a substantial depth before gradually shallowing at lower skin depth values. Our analysis hence suggests that deep deformation can occur in different tills. In high-permeability tills, deep deformation and till transport is facilitated by rapid changes in water pressure, while transport of low-permeability tills are sensitive to longer, seasonal pressure variations. Larger water-pressure magnitudes fa-cilitate deep deformation at even higher till permeabilities and slower pressure perturbations.

## Comparison to field observations

Our model demonstrates that deep slip in till occurs when remnant high water pressures at depth overcome the effective lithostatic stress gradient. The depth of slip is hence dependent on several evolving parameters including ice thickness, variability and magnitude in meltwater influx, and till properties. However, the key factor is the coupling at the ice-till interface or, in other words, the degree to which the slip speed responds to in-situ changes in the interfacial shear stress. Interestingly, the degree of resistance provided by the interfacial shear stress varies widely between glacier settings from being an almost negligible contribution to being the dominant term in the force balance (44). The coupling could also change and, in fact, is likely to change over the course of a deglaciation or a glacial cycle.

The Whillans Ice Stream provides an interesting example in this context. We conceptualize the contemporary Whillans Ice Stream as a stress-controlled setting, which is consistent with its current stick-slip behavior (43). Stress-controlled subglacial deformation results in shallow shear below the ice-bed interface (Fig. 1D and 2E). We can use Eq. 2 to estimate how a transition to speed-controlled shear would affect till transport. The basal till at this location is fine grained (45), meaning that the propagation of pressure perturbations through the system is slow (46). Observed water-pressure variations are relatively small in amplitude (7, 47), because the basal interface is not perturbed by seasonal meltwater input from the surface. A permeability estimate of $k = 4.9 \times 10^{-17} \text{ m}^2$ (48), a water-pressure amplitude of approximately 2 m head ($A_f \approx 20$ kPa) (47), and a daily forcing frequency ($f = 1.2 \times 10^{-5}$ s) does not produce deep shear as Eq.2 estimates a maximum deformation depth of $z' = 0$ m. Larger water-pressure perturbations or slower forcing frequencies are necessary for deepening deformation from hydraulic dynamics.

For the contemporary Whillans Ice Stream, our model hence predicts localized slip at the ice-till interface and relatively small till fluxes, which is consistent with observations (6, 49). Till transport is limited by the shape of the strain-profile, which in turn is dictated by the degree of non-locality in the till deformation and ploughing by protruding clasts from the ice-bed interface (46). During the last glacial maximum, however, the ice thickness in the Ross Sea was larger than today and likely characterized by greater surface slopes. Both factors increase driving stress, suggesting that lateral stresses in the margin could have played a more important role in the force balance of the ice stream than they do today, at least given comparable till properties underneath. Therefore, the ice-till interface during the last glacial maximum could have resembled speed-controlled conditions more closely. Increased normal stress at the bed also causes higher till transport (Fig. 1D), and any water-pressure variations would amplify till transport additionally (Fig. 2C). These considerations suggest that till transport was significantly larger during the last glacial maximum, which is consistent with observed bedforms on the continental shelf in the Ross Sea (50).

A transition in the coupling at the ice-till interface was observed during a previous field campaign at Black Rapids Glacier, a contemporary mountain glacier in the central Alaska Range. The till bed of Black Rapids Glacier was probed by in-situ bore-hole measurements repeatedly (9, 10). During the earlier field campaign in 1997 (10), basal hydrology and deformation did not significantly correlate with ice-surface displacement, suggesting that the subglacial bed can be approximated by speed-controlled conditions. Given estimates of permeability of $k = 2 \times 10^{-18} \text{ m}^2$, water-pressure perturbations of approximately 1 MPa, and a forcing period of 1 month (10), we estimate the maximum deformation depth to be on the order of $z' = 6$ m. This result is in broad agreement with observations that ice slip resulted from till deformation at an undetermined depth of more than 2 m below the ice-bed interface (10). The till transport in this limit is much more significant, because the entire till layer above the deep slip zone is mobilized. In contrast, later measurements at the same location on Black Rapids Glacier in 2002 (9) showed that sliding occurred partially at the ice-till interface itself and partly at shallow depth within the till layer. The glacier geometry and flow remained largely unchanged between the two field campaigns, but the recorded water-pressure variations were smaller in 2002. These seemingly contradictory observations are consistent with our model when conceptualizing Black Rapids Glacier as a largely speed-controlled setting with varying pore-water pressure amplitude that can shift the depth of basal slip away from ice-till interface and deep into the till layer itself.

Concluding, we emphasize that the motivation for providing a comparison between a highly idealized model and field observations is to demonstrate the explanatory potential of the model we have derived. There is no doubt that any actual field setting is characterized by significantly more complexity than considered here. Nonetheless, we argue that it is valuable to test how much of the observed variability can be reproduced within a relatively simple, unifying framework, where the complex coupling between ice, till and water is not confounded by additional factors such as topography, ice-ocean interactions, and hydrological processes to only name a few. A more complete understanding of the temporal evolution of basal slip at any given field site could require coupling our model to an ice-sheet model. Contrary to granular-scale DEMs, our model is conducive to model integration, because it is computationally cheap and adjustable to various grid setups.

Even in the absence of a fully coupled model, our analysis highlights the value of continuous instrumentation of field sites (9, 10, 38). Our results demonstrate the large inherent variability in subglacial deformation, which implies that measurements performed at a single point in time do not provide sufficient constraints to assess the evolution of basal slip and till transport over longer time scales. Similarly, our results highlight the importance of not reducing the dynamics of the subglacial environment only to subglacial hydrology. While subglacial hydrology is important, the effect it has on ice speed and the depth of basal slip depend sensitively on the coupling at the ice-till interface. This insight could be relevant for understanding the distinct response of glaciers to seasonal water input (42) and to climate forcing more generally.

## Methods

### Model formulation

Subglacial tills are saturated by meltwater and often have significant cohesion from clay minerals. To generalize the existing non-local, granular fluidity (NGF) model by [28] to subglacial till (original model detail listed in Supplementary Information), we allow pore pressure to perturb the stress distribution in the sediment, and add the contribution of cohesion to shear strength. The extended model is named the cohesive NGF pore fluid (CNGF-PF) model.

Typical shear-strain magnitudes in subglacial tills are $\gg 1$ [16], so we neglect the elastic contribution to shear strain ($\dot{\gamma} = \dot{\gamma}^p$). The transient evolution of pore-fluid pressure ($p_f$) is modeled through Darcian pressure diffusion [29, 30]:

$$\frac{\partial p_f}{\partial t} = \frac{1}{\phi \eta_f \beta_f} \nabla \cdot (k \nabla p_f), \qquad (3)$$

where $\eta_f$ denotes dynamic viscosity of water [Pa s], $\beta_f$ is the adiabatic fluid compressibility [Pa$^{-1}$], and $k$ is intrinsic permeability [m$^2$]. We use the effective normal stress $\sigma'_n = \sigma_n - p_f$ [Pa] instead of the dry normal stress $\sigma_n$.

Following [28], we assume that till is in the critical state throughout the domain, implying that average porosity does not change as a function of granular deformation. As a consequence, dynamic water pressure variations are negligible with respect to meltwater influx. The critical state assumption is likely justified, particularly underneath fast-moving ice [16].

We add a cohesion contribution $C$ [Pa] to the shear strength in the cooperativity length in [28]:

$$\xi(\mu) = \frac{Ad}{\sqrt{|(\mu - C/\sigma'_n) - \mu_s|}}, \qquad (4)$$

and in the local fluidity term (Eq. 5). The fluidity field is solved in the same manner as the original model [28].

### Simulation setup

We apply the model in a one-dimensional setup with simple shear (Fig. 1A) of length $L_z$. The upper normal stress (i.e. the ice weight, $\sigma_n(z = L_z)$) is constant, and normal stress linearly increases with depth due to sediment weight. In the stress-controlled experiments, the upper boundary (i.e. the ice-bed interface) applies a fixed shear stress $\tau$ on the till. In speed-controlled experiments, the upper boundary moves with a fixed shear velocity $v_x(z = L_z)$. For all simulations, the lower boundary condition for the granular phase is no slip ($v_x(L = 0) = 0$). Effective normal stress ($\sigma'_n = \sigma_n - p_f$) varies if the water pressure $p_f$ changes (Eq. 3). For the water-pressure solver, the top pressure ($p_f(z = L_z)$) is either constant or varies through time. For the experiments with variable water pressure, we apply a water-pressure forcing amplitude of $A_f = 80$ kPa that modulates effective normal stress at the top around 100 kPa. At the start of each simulation, water pressure is set to follow the hydrostatic gradient at the lower boundary ($dp_f/dz(z = 0) = \rho_f G$). Unless noted, we use the value 0.94 for the dimensionless rate-dependence parameter $b$, which is empirically constrained from laboratory experiments on glass beads [17]. We solve the model components iteratively as detailed in the Supplementary Information with specific parameter values listed in Table S1.

### Acknowledgements

We acknowledge insightful conversations with Indraneel Kasmalkar, Jason Amundson, Dougal Hansen, Lucas Zoet, Dongzhuo Li, and Alejandro Cabrales-Vargas during model development. This research was supported by the National Science Foundation through the Office of Polar Programs awards PLR-1744758 and by the U. S. Army Research Laboratory under grant W911NF-12-R0012-04.

### Author contributions

The author are listed in sequence of their contribution. A.D. developed the numerical code, performed the simulations and produced the figures; J.S. conceptualized the study; L.G. derived the analytical solution; All authors contributed to research design and to the text. The authors declare no conflict of interest.

$$g_{\text{local}}(\mu, \sigma'_n) = \begin{cases} \sqrt{d^2\sigma'_n/\rho_s}((\mu - C/\sigma'_n) - \mu_s)/(b\mu) & \text{if} \quad \mu - C/\sigma'_n > \mu_s, \quad \text{and} \\ 0 & \text{if} \quad \mu - C/\sigma'_n \le \mu_s. \end{cases} \tag{5}$$

# Supporting Information for:
# Evolving basal slip under glaciers and ice streams

### A. Damsgaard, J. Suckale and L. Goren

anders@adamsgaard.dk

## Methods

GDR-MiDi (1) presented a non-dimensional inertia number that summarizes the mechanical behavior of dry and dense granular deformation. The rate dependence on mechanical properties evolved into an empirical continuum rheology in (2) and (3), where stress and porosity depend on the inertia number. However, these models are *local*, meaning that local stresses determines the local strain-rate response alone. As a consequence, material properties do not influence shear zone width, which is not consistent with observations (1, 4–6). Granular deformation contains numerous non-local effects, where flow rates in neighboring areas influence the tendency of a sediment parcel to deform. Granular shear zones are an example of the non-locality as they have a minimum width dependent on grain characteristics (7–9).

### Original non-local granular fluidity (NGF) model

Henann and Kamrin (10) presented the non-local granular fluidity (NGF) model where a *fluidity* field variable accounts for the non-local effects on deformation. The model builds on previous continuum rheologies for granular materials (2, 3) and accurately describes non-local strain distribution in a variety of experimental settings. The physical basis for the NGF model is the statistics of a kinetic elasto-plastic mechanism, which envisions mesoscopic regions undergoing localized yielding and inducing elastic deformation in nearby, jammed regions. The consequence is a non-local picture of the flow that is characterized by a finite cooperation length. Fluidity acts as a state variable, describing the phase transition between non-deforming (jammed) and actively deforming (flowing) regions in the sediment (10, 11). All material is assumed to have a uniform porosity and to be in the critical state. The modeled sediment deforms with yield beyond the Mohr-Coulomb failure limit (10, 12), but unlike classical plastic models it includes a closed form relation that predicts the stress-strain rate relation at and slightly beyond yield.

In the NGF model, shear deformation is contributed by elastic ($\dot{\gamma}^e$) and plastic ($\dot{\gamma}^p$) shear strain rate :

$$\dot{\gamma} = \dot{\gamma}^e + \dot{\gamma}^p \tag{1}$$

where the plastic contribution to shear strain rate is given by:

$$\dot{\gamma}^p = g(\mu, \sigma_n)\mu. \tag{2}$$

Here, $\mu = \tau/\sigma_n'$ is the dimensionless ratio between shear stress ($\tau$ [Pa]) and normal stress ($\sigma_n$ [Pa]), and $g$ [s$^{-1}$] is the granular fluidity. The fluidity $g$ is a kinematic variable governed by grain velocity fluctuations and packing fraction (11), and consists of local and non-local components:

$$\nabla^2 g = \frac{1}{\xi^2(\mu)}(g - g_{\text{local}}), \tag{3}$$

The degree of non-locality is scaled by the cooperativity length $\xi$ [m], which, in turn, scales with non-local amplitude $A$ [-]:

$$\xi(\mu) = \frac{Ad}{\sqrt{|\mu - \mu_s|}}, \tag{4}$$

where $d$ [m] is the representative grain diameter and $\mu_s$ [-] is the static Coulomb yield coefficient. The local contribution to fluidity is defined as:

$$g_{local}(\mu, \sigma_n) = \begin{cases} \sqrt{d^2 \sigma_n / \rho_s}(\mu - \mu_s)/(b\mu) & \text{if } \mu > \mu_s, \text{ and} \\ 0 & \text{if } \mu \leq \mu_s. \end{cases} \tag{5}$$

where $\rho_s$ [kg m$^{-3}$] is grain mineral density, and $b$ [-] controls the non-linear rate dependence beyond yield. The failure point is principally determined by the Mohr-Coulomb constituent relation in the conditional of Eq. 5. However, the non-locality in Eq. 3 implies that deformation can occur in places that otherwise would not fail, in cases where the surrounding areas have a high local fluidity.

## Numerical solution procedure

We solve the cohesive non-local granular fluidity model with pore fluid (CNGF-PF) equations in a one-dimensional setup where simple shear occurs along a horizontal axis $x$, orthogonal to a vertical axis $z$. The spatial domain is $L_z = 8$ m long and is discretized into cells with equal size to the representative grain size $d$. The upper boundary, i.e. the "ice-bed interface", exerts effective normal stress and shear stress on the granular assemblage. We neglect the minuscule contribution to material shear strength from water viscosity. The effective normal stress within the layer is found by adding the lithostatic contribution that increases with depth to the normal stress applied from the top:

$$\sigma_n(z) = \sigma_{n,top} + (1-\phi)\rho_s G(L_z - z), \tag{6}$$

where $G$ [m s$^{-2}$] is gravitational acceleration, and

$$\sigma_n'(z) = \sigma_n(z) - p_f(z). \tag{7}$$

Normal stress $\sigma_n(z = L_z)$ and fluid pressure $p_f(z = L_z)$ at the top are described by the boundary condition as constant or time-variable values. We compute the apparent friction coefficient $\mu$ as:

$$\mu(z) = \mu_{0,top} \frac{\sigma_{n,top}'}{\sigma_n'(z)}. \tag{8}$$

where $\mu_{0,top}$ is the initial friction at the top at $t = 0$. The shear stress $\tau(z) = \mu(z)\sigma_n(z)$ is constant in time and space for stress-controlled experiments, and dynamic for speed-controlled experiments.

We assign depth coordinates $z_i$, granular fluidity $g_i$, and fluid pressure $p_{f,i}$ to a regular grid with ghost nodes and cell spacing $\Delta z$. The ghost nodes are imaginary grid nodes outside of the top and bottom boundaries, and their values are dynamically adjusted to provide the desired boundary condition. The fluidity field $g$ is solved for a set of mechanical forcings ($\mu$, $\sigma_n'$, boundary conditions for $g$), and material parameters ($A$, $b$, $d$). We rearrange Eq. 3 and split the Laplace operator ($\nabla^2$) into a 1D central finite difference 3-point stencil. We apply an iterative scheme to relax the following equation at each grid node $i$:

$$g_i = (1 + \alpha_i)^{-1}\left(\alpha_i g_{local}(\sigma_{n,i}', \mu_i) + \frac{g_{i+1} + g_{i-1}}{2}\right), \tag{9}$$

where

$$\alpha_i = \frac{\Delta z^2}{2\xi^2(\mu_i)}. \tag{10}$$

We apply fixed-value (Dirichlet) boundary conditions for the fluidity field ($g(z = 0) = g(z = L_z) = 0$). This condition is appropriate for confined flows. Neumann boundary conditions, which are not used here, create a velocity profile resembling a free surface flow.

The pore-pressure solution (Eq. 3 in the main text) is constrained by a hydrostatic pressure gradient at the bottom ($dp_f/dz(z = 0) = \rho_f G$), and a pressure forcing at the top, for example sinusoidal: $p_f(z = L_z) = A_f \sin(2\pi f t) + p_{f,0}$. Here, $A_f$ is the forcing amplitude [Pa], $f$ is the forcing frequency [1/s], and $p_{f,0}$ is the mean pore pressure over time [Pa]. As for the granular fluidity field (Eq. 3), we also use operator splitting and finite differences to solve the equation for pore-pressure diffusion (Eq. 3 in the main text):

$$\Delta p_{f,i} = \frac{1}{\phi_i \eta_f \beta_f} \frac{\Delta t}{\Delta z}\left(\frac{2k_{i+1}k_i}{k_{i+1} + k_i}\frac{p_{i+1} - p_i}{\Delta z} - \frac{2k_i k_{i-1}}{k_i + k_{i-1}}\frac{p_i - p_{i-1}}{\Delta z}\right). \tag{11}$$

For each time step $\Delta t$, we compute a solution to Eq. 11 through the Crank-Nicholson method (13–15). In this procedure, the pressure field at $t + \Delta t$ is found by mixing explicit and implicit solutions with equal weight. The method is unconditionally stable and second-order accurate in time and space. Our implementation of grain and fluid dynamics is highly efficient, and for the presented experiments each time step completes in less than 1 ms on a single CPU core.

## Speed-controlled experiments

The model form presented above is suited for resolving strain rate and shear velocity from a given stress forcing, i.e., in a stress-controlled setup. However, the basal conditions under glaciers and ice streams are highly variable and certain cases are better approximated by a speed-controlled limit where a specified shear speed at the interface results in a strain-rate distribution and shear stress inside the subglacial bed. For our system of equations, this case represents an inverse problem that we compute by adjusting the applied friction at the top until the resultant shear speed matches the desired value. We implement an automatic iterative procedure that can be set to match a shear speed, or limit the speed to a specified value. First, we calculate an initial top speed value $v_x^*$ is calculated in a forward manner on the base of an arbitrary value for friction $\mu^*$. We use the difference between $v_x^*$ and the desired speed $v_x^d$ in the calculation of a normalized residual $r$:

$$r = \frac{v_x^d - v_x^*}{v_x^* + 10^{-12}}. \tag{12}$$

We add the value $10^{-12}$ to the denominator to avoid division by zero if the initial applied friction value $\mu^*$ does not cause yield. If the residual value $r$ is negative, the current applied friction produces a shear speed that exceeds the desired value, and vice versa. If the absolute value of the residual exceeds the tolerance criteria ($|r| > 10^{-3}$), we adjust the applied friction:

$$\mu_{new}^* = \mu^*(1.0 + \theta r), \tag{13}$$

where $\theta = 10^{-2}$ is a chosen relaxation factor. The computations are then rerun with the new applied friction until the tolerance criteria is met.

The grain-water model is written in C and is available under free-software licensing at https://src.adamsgaard.dk/1d_fd_simple_shear. All simulation parameters can be specified as command-line arguments. Run ./1d_fd_simple_shear -h after compiling for usage information. The results and figures in this paper can be reproduced by following the instructions in the experiment repository for this publication, available at https://src.adamsgaard.dk/manus_continuum_granular1_exp.

## Analytical solution for maximum deformation depth

Under the assumption that the bed is a semi-infinite halfspace, we can solve for the depth profile and transient behavior of effective normal stress $\sigma_n'$ analytically by extending a solution for dispersion of a sinusoidal forcing through a diffusive medium (16). Here, $z'$ is the depth below the ice-bed interface, i.e., $z' = L_z - z$:

$$\sigma_n'(z', t) = \sigma_n + (\rho_s - \rho_f)Gz' - p_{f,top} - A_f \exp\left(-\frac{z'}{d_s}\right)\sin\left(\omega t - \frac{z'}{d_s}\right). \tag{14}$$

We compute the vertical gradient of the effective normal stress by evaluating,

$$\frac{d\sigma_n'}{dz'}(z', t) = (\rho_s - \rho_f)G + \frac{A_f}{d_s}\exp\left(-\frac{z'}{d_s}\right)\left[\sin\left(\omega t - \frac{z'}{d_s}\right) + \cos\left(\omega t - \frac{z'}{d_s}\right)\right], \tag{15}$$

where $\omega = 2\pi f$ is the circular forcing frequency [s$^{-1}$]. For this study, we want to find the depth $z'$ where $d\sigma_n'/dz' = 0$. At this depth the effective normal stress is at a minimum and deep deformation can occur. In our simulations we observe that the deepest deformation occurs when water pressure is at its minimum at the ice-bed interface, which means that $t = 3\pi/2\omega$:

$$0 = \sin\left(\frac{3\pi}{2} - \frac{z'}{d_s}\right) + \cos\left(\frac{3\pi}{2} - \frac{z'}{d_s}\right) + \frac{(\rho_s - \rho_f)Gd_s}{A_f}\exp\left(\frac{z'}{d_s}\right) \tag{16}$$

In the main text the above equation is presented in shorter form using the identity, $\sin(x) + \cos(x) = \sqrt{2}\sin(x + \pi/4)$. With a sinusoidal water-pressure forcing, the above equation has no solution if $d\sigma_n'/dz(z = 0) > 0$. This is the case if the pressure perturbation is too weak to reverse the effective normal stress curve at depth, causing shear deformation to occur at the top throughout the water-pressure cycle.

We use Brent's method (15) for numerically finding depth ($z'$) values that satisfy the above equation within $z' \in [0; 5d_s]$. Our implementation, the program `max_depth_simple_shear`, takes command-line arguments of the same format as the main NGF program, `1d_fd_simple_shear`, and prints the maximum deformation depth ($z'$) as the first column of output, and the skin depth ($d_s$) as the second column. See "`max_depth_simple_shear -h`" for usage details.

## Parameter sensitivity test

Figure S3 contains a systematic analysis of parameter influence in the model equations. Several observations emerge from this parameter sensitivity analysis. The representative grain size $d$ has a major influence on the strain distribution, where finer materials show deeper deformation. The material is slightly weaker with larger grain sizes. The shear zone is more narrow with higher material static friction coefficients ($\mu_s$), as the material is less prone to failure. Our implementation of cohesion does not influence strain after yield. Static friction and cohesion both linearly scale the bulk friction, as expected with Mohr-Coulomb materials. The non-local amplitude $A$ slightly changes the curvature of the shear strain profile, but does not affect the overall friction. There is a significant strengthening when the bed thickness $L_z$ begins to constrict the shear zone thickness.

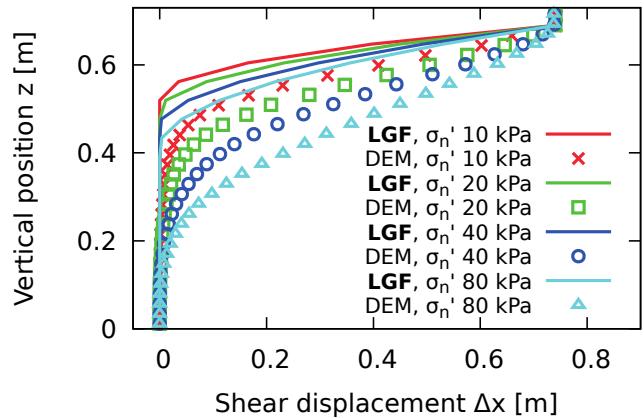

**Fig. S1.** Strain distribution in local-only granular fluidity (LGF) model ($A \approx 0$).

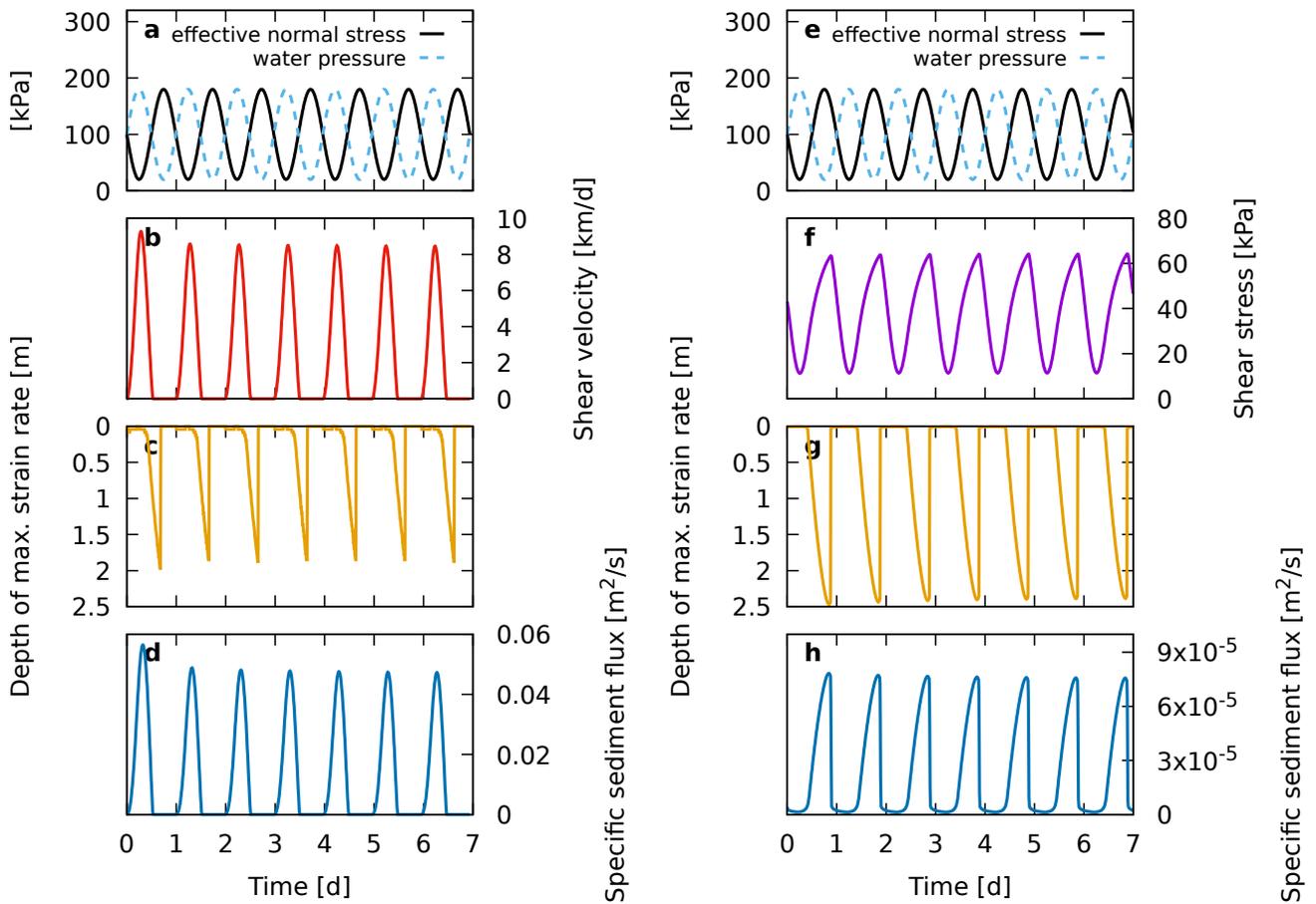

**Fig. S2.** Deformation dynamics during sinusoidal water-pressure forcing from the top. Stress and shear velocity are measured at the top of the sediment bed. a-d) Stress-controlled setup with applied shear stress $0.4\sigma'_n$. e-h) speed-controlled setup with applied shear velocity $v_x = 1$ km/a.

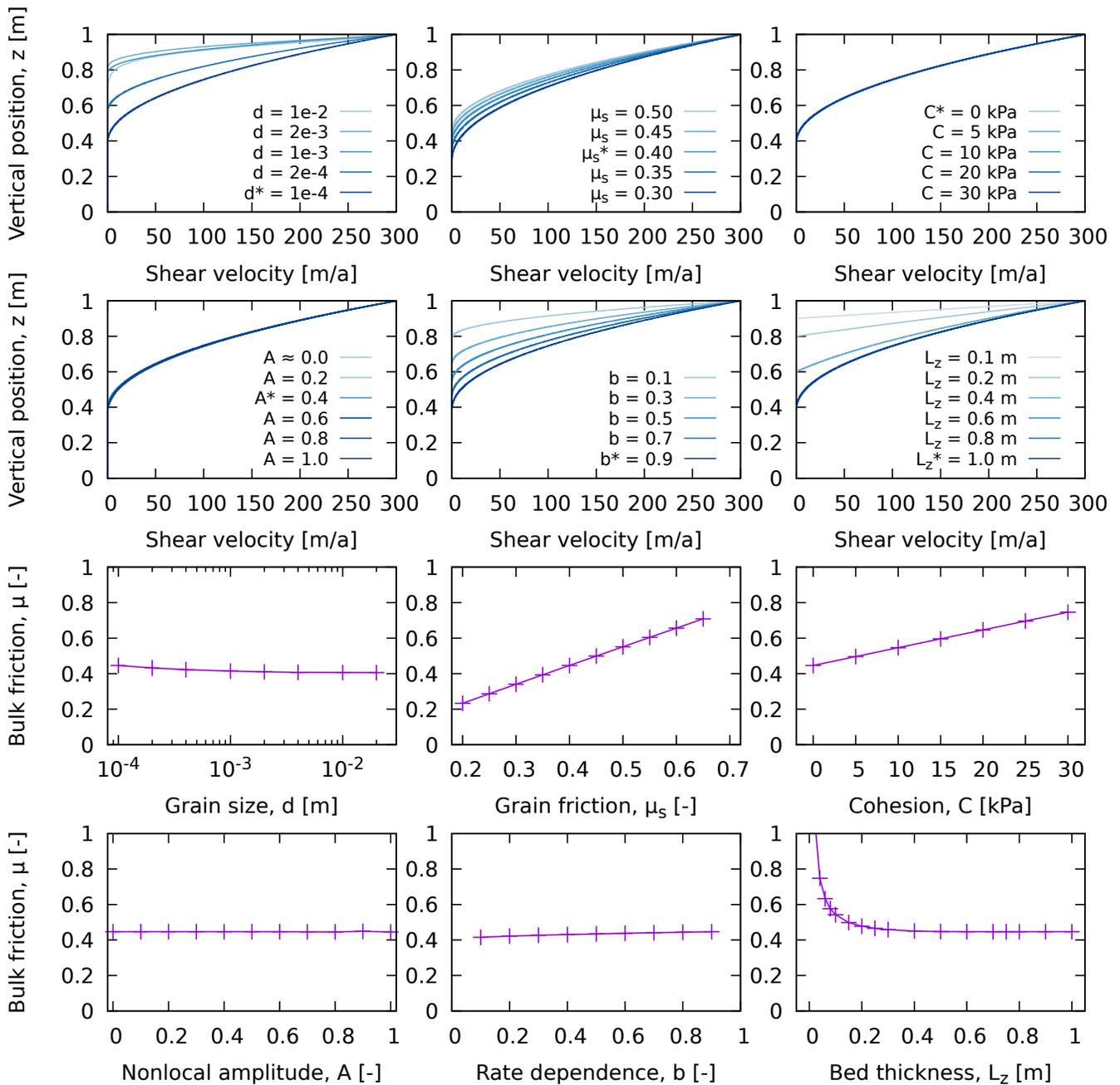

**Fig. S3.** Analysis of parameter influence on steady-state strain distribution and bulk friction during shear. All experiments are performed under constant shear rate of 300 m a$^{-1}$ and a normal stress of $\sigma'_n = 100$ kPa. Parameter values marked with an asterisk (*) are used outside of the individual parameter sensitivity tests.

**Table S1.** Material parameters for model simulations emulating discrete element method (DEM) particles (17), and two idealized tills with varying grain size distribution. Parameter values from the literature are used where marked with a reference symbol: a: (17), b: (18), c: (12), d: (19), e: (20).

| Parameter | Symbol | Units | DEM particles | Idealized till |
|---|---|---|---|---|
| Friction coefficient | $\mu_s$ | – | $0.404_a$ | $0.40$ |
| Cohesion | $C$ | kPa | $0_a$ | $0$ |
| Representative grain size | $d$ | m | $0.04$ | $1.0 \times 10^{-3}$ |
| Hydraulic permeability | $k$ | m$^2$ | $2 \times 10^{-17}_b$ | $2 \times 10^{-17}$ |
| Nonlocal amplitude | $A$ | – | $0.50$ | $0.40_c$ |
| Nonlinear rate dependence | $b$ | – | $0.022$ | $0.94_c$ |
| Grain material density | $\rho_s$ | kg m$^{-3}$ | $2.6 \times 10^3_a$ | $2.6 \times 10^3$ |
| Porosity | $\phi$ | – | $0.25_a$ | $0.25$ |
| Dynamic fluid viscosity | $\eta_f$ | Pa s | $1.787 \times 10^{-3}_d$ | $1.787 \times 10^{-3}$ |
| Adiabatic fluid compressibility | $\beta_f$ | Pa$^{-1}$ | $3.9 \times 10^{-10}_e$ | $3.9 \times 10^{-10}_e$ |